\documentclass[pdflatex,sn-standardnature,iicol]{sn-jnl}
\jyear{2022}%
\usepackage{tablefootnote}
\usepackage{siunitx}
\ifdefined\qty\else
  \ifdefined\NewCommandCopy
    \NewCommandCopy\qty\SI
  \else
    \NewDocumentCommand\qty{O{}mm}{\SI[#1]{#2}{#3}}
  \fi
\fi
\ifdefined\unit\else
  \ifdefined\NewCommandCopy
    \NewCommandCopy\unit\si
  \else
    \NewDocumentCommand\unit{O{}m}{\si[#1]{#2}}
  \fi
\fi

\usepackage{caption}
\usepackage{subcaption}
\unnumbered
\usepackage{placeins}

\usepackage{multibib}

\begin{document}

\title[229Th rad. dec.]{Observation of the radiative decay of the ${}^{229}\mathrm{Th}$ nuclear clock isomer}

\author*[1,12]{\fnm{Sandro} \sur{Kraemer}}\email{sandro.kraemer@kuleuven.be}

\author[2]{\fnm{Janni} \sur{Moens}}

\author[3,1]{\fnm{Michail} \sur{Athanasakis-Kaklamanakis}}

\author[1]{\fnm{Silvia} \sur{Bara}}

\author[4]{\fnm{Kjeld} \sur{Beeks}}

\author[1]{\fnm{Premaditya} \sur{Chhetri}}

\author[3]{\fnm{Katerina} \sur{Chrysalidis}}

\author[1]{\fnm{Arno} \sur{Claessens}}

\author[1]{\fnm{Thomas E.} \sur{Cocolios}}

\author[5]{\fnm{Jo\~ao M.} \sur{Correia}}

\author[1]{\fnm{Hilde} \sur{De Witte}}

\author[1]{\fnm{Rafael} \sur{Ferrer}}

\author[1]{\fnm{Sarina} \sur{Geldhof}}

\author[3]{\fnm{Reinhard} \sur{Heinke}}

\author[4]{\fnm{Niyusha} \sur{Hosseini}}

\author[1]{\fnm{Mark} \sur{Huyse}}

\author[6]{\fnm{Ulli} \sur{Köster}}

\author[1]{\fnm{Yuri} \sur{Kudryavtsev}}

\author[7,8,9]{\fnm{Mustapha} \sur{Laatiaoui}}

\author[3,10]{\fnm{Razvan} \sur{Lica}}

\author[2]{\fnm{Goele} \sur{Magchiels}}

\author[1]{\fnm{Vladimir} \sur{Manea}}

\author[11]{\fnm{Clement} \sur{Merckling${}^{11}$}}

\author[2]{\fnm{Lino} \sur{M. C. Pereira}}

\author[8,9]{\fnm{Sebastian} \sur{Raeder}}

\author[4]{\fnm{Thorsten} \sur{Schumm}}

\author[1]{\fnm{Simon} \sur{Sels}}

\author[12]{\fnm{Peter G.} \sur{Thirolf}}

\author[2]{\fnm{Shandirai Malven} \sur{Tunhuma}}

\author[1]{\fnm{Paul} \sur{Van Den Bergh}}

\author[1]{\fnm{Piet} \sur{Van Duppen}}

\author[2]{\fnm{André} \sur{Vantomme}}

\author[1]{\fnm{Matthias} \sur{Verlinde}}

\author[2]{\fnm{Renan} \sur{Villarreal}}

\author[5]{\fnm{Ulrich} \sur{Wahl}}

\affil[1]{\orgname{KU Leuven}, \orgdiv{Kern- en Stralingsfysica}, \orgaddress{\street{Celestijnenlaan 200D}, \postcode{3001} \city{Leuven}, \country{Belgium}}}

\affil[2]{\orgname{KU Leuven}, \orgdiv{Quantum Solid State Physics}, \orgaddress{\street{Celestijnenlaan 200D}, \postcode{3001} \city{Leuven}, \country{Belgium}}}

\affil[3]{\orgname{CERN}, \orgaddress{\street{Esplanade des Particules 1}, \postcode{1211} \city{Genève}, \country{Switzerland}}}

\affil[4]{\orgdiv{Atominstitut}, \orgname{TU Wien}, \orgaddress{\street{Stadionallee 2}, \postcode{1020} \city{Wien}, \country{Austria}}}

\affil[5]{\orgdiv{Centro de Ciências e Tecnologias Nucleares, Departamento de Engenharia e Ciências Nucleares, Instituto Superior Técnico}, \orgname{Universidade de Lisboa}, \orgaddress{\street{Estrada Nacional 10, ao km 139,7}, \postcode{2695-066} \city{Bobadela}, \country{Portugal}}}

\affil[6]{\orgname{Institut Laue-Langevin}, \orgaddress{\street{71 avenue des Martyrs}, \postcode{3804} \city{Grenoble CEDEX}, \country{France}}}

\affil[7]{\orgdiv{Department Chemie}, \orgname{Johannes-Gutenberg-Universität}, \orgaddress{\street{Fritz-Straßmann-Weg 2}, \postcode{55128} \city{Mainz}, \country{Germany}}}

\affil[8]{\orgname{Helmholtz Institut Mainz}, \orgaddress{\street{Staudingerweg 18}, \postcode{55128} \city{Mainz}, \country{Germany}}}

\affil[9]{\orgname{GSI Helmholtzzentrum für Schwerionenforschung}, \orgaddress{\street{Planckstraße 1}, \postcode{64291} \city{Darmstadt}, \country{Germany}}}

\affil[10]{\orgname{Horia Hulubei National Institute of Physics and Nuclear Engineering}, \orgaddress{\street{30 Reactorului}, \postcode{077125} \city{Bucharest}, \country{Romania}}}

\affil[11]{\orgname{Imec}, \orgaddress{\street{Kapeldreef 75}, \postcode{3001} \city{Leuven}, \country{Belgium}}}

\affil[12]{\orgname{Ludwig-Maximilians-Universität München}, \orgaddress{\street{Am Coulombwall 1}, \postcode{85748} \city{Garching}, \country{Germany}}}

\keywords{thorium, 229Th, isomer, vacuum-ultraviolet, radiative decay, optical clock, metrology}

\maketitle

\renewcommand{\figurename}{Figure}
\renewcommand{\tablename}{Table}

\noindent
\textbf{
\noindent
The nucleus of the radioisotope thorium-229 (${}^{229}$Th) features an isomer with an exceptionally low excitation energy that enables direct laser manipulation of nuclear states. For this reason, it is a leading candidate for use in next-generation optical clocks \cite{peik_nuclear_2003,beeks_thorium-229_2021}. 
This nuclear clock will be a unique tool, amongst others, for tests of fundamental physics \cite{Peik_2021,thirolf_improving_2019}. 
While first indirect experimental evidence for the existence of such an extraordinary nuclear state is significantly older \cite{kroger_features_1976}, the proof of existence has been delivered only recently by observing the isomer's electron conversion decay \cite{von_der_wense_direct_2016} and its hyperfine structure in a laser spectroscopy study \cite{thielking_laser_2018}, revealing information on the isomer's excitation energy, nuclear spin and electromagnetic moments. Further studies reported the electron conversion lifetime and refined the isomer's energy \cite{seiferle_lifetime_2017,seiferle_energy_2019,sikorsky_measurement_2020}. 
In spite of recent progress, the isomer's radiative decay, a key ingredient for the development of a nuclear clock, remained unobserved.}

\textbf{
In this Letter, we report the detection of the radiative decay of this low-energy isomer in thorium-229 (${}^{229\mathrm{m}}$Th). By performing vacuum-ultraviolet spectroscopy of ${}^{229\mathrm{m}}$Th incorporated into large-bandgap CaF$_{2}$ and MgF$_{2}$ crystals at the ISOLDE facility at CERN, the photon vacuum wavelength of the isomer's decay is measured as \qty{148.71(42)}{\nano\meter}, corresponding to an excitation energy of \qty{8.338(24)}{\electronvolt}. 
This value is in agreement with recent measurements \cite{seiferle_energy_2019, sikorsky_measurement_2020}, and decreases the uncertainty by a factor of seven. The half-life of ${}^{229\mathrm{m}}$Th embedded in MgF$_{2}$ is determined to be \qty{670(102)}{\second}. The observation of the radiative decay in a large-bandgap crystal has important consequences for the design of a future nuclear clock and the improved uncertainty of the energy eases the search for direct laser excitation of the atomic nucleus.}

The isotope ${}^{229}$Th and its low-energy isomer have inspired research for decades and the prospect of developing an optical clock using a nuclear transition has intensified efforts \cite{Peik_2021}. 
A particular focus lies on the precise measurement of properties relevant for an optical clock involving the direct laser manipulation of nuclear states. Values of the isomer's excitation energy reported in the literature have changed significantly over time \cite{von_der_wense_229th_2020}. 
Recent measurements using conversion electron spectroscopy of electrically neutral thorium-229 atoms (${}^{229}$Th${}^0$) resulted in an excitation energy of \qty{8.28(17)}{\electronvolt}, corresponding, for radiative decay, to photons of a vacuum wavelength of \qty{149.7(31)}{\nano\meter} \cite{seiferle_energy_2019}. 
A measurement of gamma-ray energy differences of nuclear transitions feeding the isomer and the ground state, using a magnetic micro-calorimeter, revealed an energy of \qty{8.10(17)}{\electronvolt}, with an associated vacuum wavelength of \qty{153.1(32)}{\nano\meter} \cite{sikorsky_measurement_2020}.

A half-life of \qty{7(1)}{\micro \second} has been reported for ${}^{229\mathrm{m}}$Th${}^0$ deposited onto a microchannel plate detector, an environment in which electron conversion is expected to constitute the dominant decay path \cite{seiferle_lifetime_2017}. 
Theoretical estimates for the radiative decay half-life vary by an order of magnitude between about \num{e3} and \qty{e4}{\second} \cite{tkalya_radiative_2015, minkov_reduced_2017, ruchowska_nuclear_2006}. 
For a dominating radiative decay, which is required for the clock application, the non-radiative decay channels, e.g. via electron conversion decay, need to be sufficiently suppressed. The latter requires charged ${}^{229\mathrm{m}}$Th ions to have an electron binding energy larger than the isomer's decay energy.

Currently, two different routes towards a nuclear clock are pursued. These consist of an approach with triply-charged thorium ions stored in a radio-frequency ion trap and a solid-state approach with a thorium-doped large-bandgap crystal \cite{beeks_thorium-229_2021,von_der_wense_229th_2020}. 
For the latter, theoretical studies suggest that, at specific lattice positions, the conversion-electron decay channel of the isomer is suppressed and the radiative decay channel becomes dominant \cite{dessovic_229_2014,pimon_dft_2020}.
Despite various attempts, the observation of the radiative decay of ${}^{229\mathrm{m}}$Th following the radioactive alpha ($\alpha$) decay of ${}^{233}$U or the excitation of the nucleus by e.g. synchrotron light  has hitherto not been conclusive \cite{stellmer_toward_2018,stellmer_radioluminescence_2015,utter_reexamination_1999,masuda_x-ray_2019}. 

In this work, an alternative approach to produce the $^{229}$Th isomer\cite{verlinde_alternative_2019} and thus overcome limitations of previous attempts to detect and characterise its radiative decay is successfully pursued. The isomeric state is populated in the beta ($\beta$-) decay of ${}^{229}$Ac with a half-life of 62.7(5)\si{\minute}\cite{browne_nuclear_2008} incorporated in a large-bandgap crystal. This increases the total feeding probability of the isomer by a factor between 7 and 47 compared to the $\alpha$ decay of ${}^{233}$U \cite{verlinde_alternative_2019}. The ${}^{229}$Ac sample is obtained from the $\beta$-decay chain of ${}^{229}$Fr and ${}^{229}$Ra ion beams implanted in CaF$_{2}$ and MgF$_{2}$ crystals at \qty{30}{\kilo\electronvolt}. Vacuum-ultraviolet spectroscopy is performed to study the photon spectrum emitted from the crystals under favourable radioluminescence background conditions, owing to the approach of populating the isomer via beta decay \cite{stellmer_radioluminescence_2015,stellmer_feasibility_2016}. In addition, emission channeling \cite{hofsass_emission_1991,wahl_position-sensitive_2004,silva_versatile_2013} is used to determine the lattice sites occupied by thorium after implantation of francium and radium, which are expected to play a crucial role in suppressing the conversion-electron decay channel \cite{dessovic_229_2014,pimon_dft_2020}. 

\subsection{Radioactive ion beam production}

Radioisotopes are produced at the ISOLDE facility at CERN \cite{borge_isolde_2017}
in a nuclear reaction by impinging a \qty{1.4}{\giga\electronvolt} proton beam on a uranium carbide target heated to about \qty{2000}{\degreeCelsius}. 
The nuclear-reaction products diffuse through the target and effuse towards the ion source, where they are surface-ionized to a singly-charged state, subsequently accelerated to \qty{30}{\kilo\electronvolt}, separated according to their mass-over-charge ratio using a dipole magnet and transported to the experimental setup. The 
radioactive ion beam of mass number $A=229$ is implanted at room temperature into large-bandgap, vacuum-ultraviolet-transparent CaF$_{2}$ ($E_\mathrm{gap}$ = \qty{12.1}{eV} \cite{Rubloff1971}) and MgF$_{2}$ (\qty{12.4}{eV} \cite{Thomas1973}) crystals. This results, for CaF$_{2}$ (MgF$_{2}$), in an implantation depth distribution with a projected range of 17.0 (16.0) \qty{}{\nano\meter} and a straggling of 3.5 (2.7) \qty{} {\nano\meter}, respectively \cite{ZIEGLER20101818}. 
A summary of the properties of the radioactive ion beams and the large-bandgap crystals used for implantation are given in the Methods section. 

\subsection{Vacuum-ultraviolet spectroscopy}

The radioactive beam is implanted into different crystals positioned on a sample holder wheel in the vacuum-ultraviolet (VUV) spectroscopy setup. High-purity germanium and lanthanum-bromide gamma ($\gamma$) radiation detectors are placed around the implantation position and used to deduce the beam production rates (see Methods section) and to determine the total amount of implanted radioisotopes in the crystal. 
The wheel is rotated after an implantation time between \qty{45}{\minute} and \qty{2}{\hour}, placing the crystal in front of the entrance slit of a VUV spectrometer (Resonance Ltd. VM180). The distance between the implanted surface of the crystal and the entrance slit is about \qty{3}{\milli\meter}. 
The spectrometer has a large numerical aperture of F/1.2 and houses a plane grating Czerny-Turner mount configuration with off-axis parabolic mirrors. A Hamamatsu R8487 photomultiplier tube detects photons in the wavelength range between \num{115} and \qty{195}{\nano\meter}. 
VUV spectra of photons created in the crystals are recorded by rotating the grating, thereby scanning the wavelength and recording the photon count rate in the photomultiplier detector.
The instrument operates in the first diffraction order and is calibrated using external calibration light sources (see Methods section). 
The efficiency of photon detection for an isotropically emitting point-like source positioned at \qty{3}{\milli\meter} from a \qty{3}{\milli\meter} wide slit yields about \qty{0.1}{\percent} at \qty{150}{\nano\meter} (see Methods section).

\begin{figure*}[hbt!]
	\centering
	\includegraphics[width=0.9\textwidth]{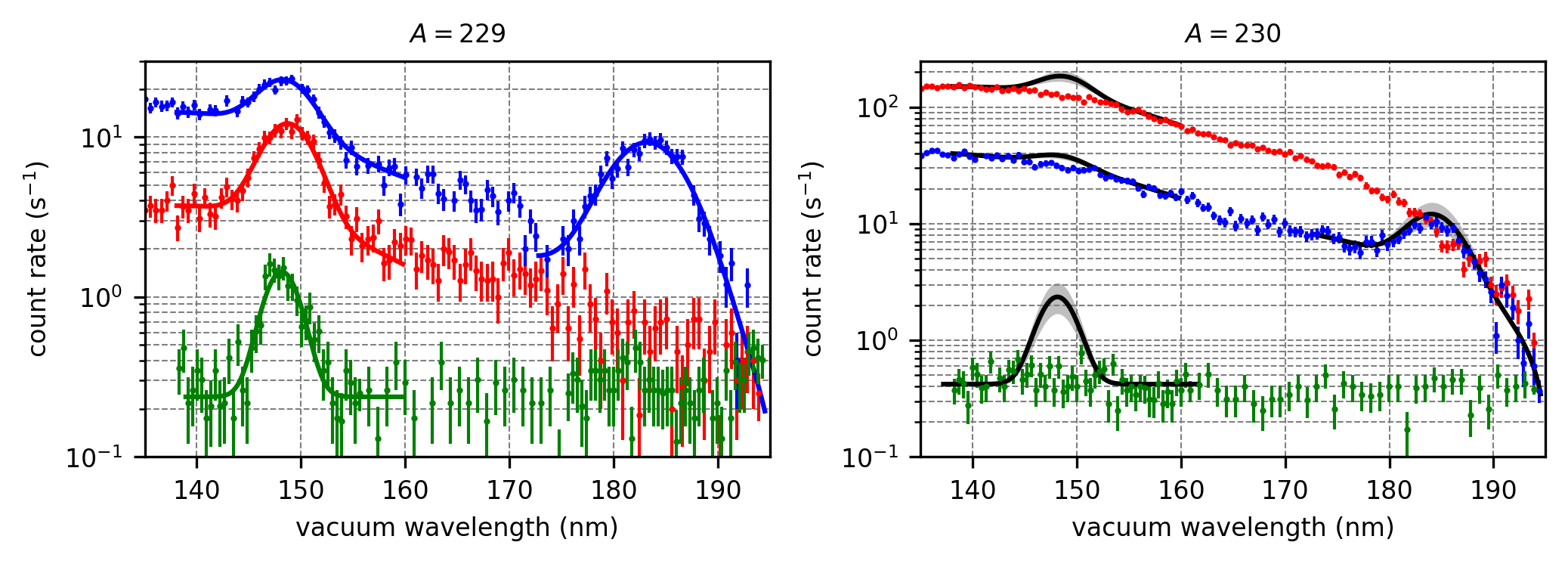}
	
	\caption{VUV spectra recorded with a \qty{3}{\milli\meter} entrance slit after implantation of $A=229$ (left panel)  and $A=230$ (right panel) beams in a  
	MgF$_2$ - \qty{5}{\milli\meter} (red), a CaF${}_2$ - \qty{5}{\milli\meter} (blue) and a CaF${}_2$ - \qty{50}{\nano\meter} (green) crystal. Wavelength data points are recorded for \qty{11}{\second} per grating setting for the \qty{5}{mm} thick crystals whereas for the \qty{50}{\nano\meter} crystal \qty{23}{s} and  \qty{35}{s}  per grating setting are used for the $A=229$ and $A=230$ data, respectively. The drop in intensity in the $A=229$ spectrum with the CaF${}_2$ - \qty{5}{\milli\meter} at \qty{172}{\nano\meter} coincides with an outage of the grating control causing a delay in the scanning procedure and resulting in a change of background activity. The solid red, blue and green lines show a fit to the data for the $A=229$ data (see Methods section). Based on a scaling with the implanted activity and assuming that the observed line at \qty{183}{\nano\meter} in the $A=229$ spectra is due to crystal defects, the black solid line represents the expected signal in the $A=230$ spectra. The grey uncertainty band includes the uncertainty on the radioactive ion beam production rates (see Methods section).} 
	\label{fig:A229A230}
\end{figure*}

The radiative decay of ${}^{229\mathrm{m}}$Th is expected to manifest itself as monochromatic photon emission resulting in a narrow peak around \qty{150}{\nano\meter} in the wavelength spectrum \cite{seiferle_energy_2019, sikorsky_measurement_2020}. 
To observe this signal, VUV spectroscopy is performed on a series of CaF${}_2$ and MgF${}_2$ crystals doped with $A=229$ and $A=230$ beams. The $A=230$ beam implantation is used as a reference measurement, as the ${}^{229}$Th isomer is not populated but where conditions regarding beam intensity and radioluminescence background are similar to those obtained with $A=229$ beams. 
Typical VUV spectra are shown in Fig.~\ref{fig:A229A230}. For both types of crystal, MgF${}_2$ and CaF${}_2$, a peak at \qty{148.7}{\nano\meter} on top of a continuous background is observed in all spectra obtained with the $A=229$ implanted beam. This \qty{148.7}{\nano\meter} peak is absent in the spectra taken with the $A=230$ beam. 
Additionally, for both the $A=229$ and $A=230$ implantations a peak at \qty{183}{\nano\meter} is observed for the \qty{5}{\milli\meter} thick CaF${}_2$ crystal. The continuous background stems from Cherenkov photon emission induced by the $\beta$-decay chain of the implanted radioactive isotopes \cite{stellmer_feasibility_2016,browne_nuclear_2008} and the constant detector dark count rate of typically less than \qty{1}{\Hz}. 
The VUV photon peak can be caused by the radiation-induced  excitation of electronic modes associated with crystal defects \cite{beeks2022nuclear}.
The latter should be observed for both $A=229$ and $A=230$ spectra. Assuming that the two VUV photon peaks in the $A=229$ spectra stem from crystal defects, the expected signal for the $A=230$ case can be calculated considering the peak intensities in the $A=229$ spectra and the different instantaneous radiation from the radioactive decays (see  black lines in Fig.~\ref{fig:A229A230}).
The agreement of the observed intensity for the \qty{183}{\nano\meter} peak in the $A=229$ and $A=230$ spectra associates this structure with a CaF${}_2$ crystal defect.
The observation of the \qty{148.7}{\nano\meter} peak in the $A=229$ spectra and its absence in $A=230$ spectra suggests that these photons are due to the radiative decay of ${}^{229\mathrm{m}}$Th.

The time evolution of the intensity of the \qty{148.7}{\nano\meter} photon peak is studied by implanting the $A=229$ beam in MgF${}_2$. After one hour of implantation, the source is moved in front of the \qty{2}{\milli\meter} entrance slit of the VUV spectrometer and consecutive scans are acquired around the \qty{148.7}{\nano\meter} peak. 
From the obtained spectra, the \qty{148.7}{\nano\meter} peak and background amplitudes at the peak position are deduced and plotted as a function of time after the end of the implantation period (see Fig.~\ref{fig:lifetime}). 
The time evolution of the peak amplitude, which is significantly different from its precursor's beta decay activity, is compatible with a mother ${}^{229}$Ac - daughter ${}^{229\mathrm{m}}$Th decay sequence. 
To deduce the ${}^{229\mathrm{m}}$Th half life from these data, the activities of the full decay chain of ${}^{229}$Fr, ${}^{229}$Ra, ${}^{229}$Ac and ${}^{229\mathrm{m}}$Th are calculated using the half-lives of the first three isotopes known from the literature and the measured beam intensity of ${}^{229}$Fr and ${}^{229}$Ra during the implantation period. By scaling the calculated activity to take into account the total VUV spectometer efficiency, a half-life value of \qty{670(102)}{\second} for the decay of ${}^{229\mathrm{m}}$Th embedded in a MgF${}_2$ crystal is obtained.

To better constrain the \qty{148.7}{\nano\meter} peak position, scans using small slit sizes are performed. The latter increase the resolution and minimize the  influence of the distribution of the implanted radioactive beam in the crystal on the energy uncertainty, at the cost of a decreased efficiency.  A typical wavelength scan at \qty{0.5}{\milli\meter} slit size is shown in Fig.~\ref{fig:narrowspectrum} (see Methods section) and the obtained wavelength values from single scans in different crystals and at different slit sizes are shown in Fig.~\ref{fig:energy}. 

For implantations in MgF${}_2$, a mean wavelength of \qty{148.688(66)}{\nano\meter} is obtained, while \qty{5}{\milli\meter} thick and \qty{50}{\nano\meter} thin CaF${}_2$ yield \qty{148.75(13)}{\nano\meter} and \qty{148.623(77)}{\nano\meter}, respectively. The stated uncertainties represent only statistical uncertainties from a $\chi^2$-optimization and do not include systematic effects from the calibration procedure.

The observation of the peak at \qty{148.7}{\nano\meter} in both the MgF$_2$ and CaF$_2$ crystals with wavelength values within the statistical uncertainty (1 $\sigma$ confidence level), the absence of the signal using the $A=230$ decay chain and the time behaviour of the signal's amplitude, leads to the conclusion that the \qty{148.7}{\nano\meter} photons stem from the radiative decay of the ${}^{229}$Th isomer.

For the final wavelength value only measurements with slit settings $\leq$ 0.5~mm are used in order to limit the systematic uncertainty. The latter accounts for calibration shifts due to different positioning of the wavelength calibration light source and the crystal sample, the distribution of the radioactive source in front of the entrance slit and the reproducibility of the grating position. This leads to a conservative systematic uncertainty of 0.41~nm (see Methods section).

This results in an average wavelength value of 
\begin{equation*}
	148.71 \pm0.06\mathrm{(stat.)} \pm 0.41 \mathrm{(syst.)}~\qty{}{\nano\meter}
\end{equation*}
corresponding to an excitation energy of the isomer of
\begin{equation*}
	8.338 \pm0.003\mathrm{(stat.)} \pm 0.023 \mathrm{(syst.)}~\mathrm{eV}.
\end{equation*}

From the observed rate of the \qty{148.7}{\nano\meter} photons, a lower limit of the fraction of ${}^{229\mathrm{m}}$Th that decays by photon emission can be determined. Using the calculated efficiency of the VUV spectrometer, the deduced production rate of ${}^{229}$Fr and ${}^{229}$Ra, and an estimate of the crystal transmission, a lower limit between 1 and 7 $\%$ is obtained assuming a total beta feeding probability of the isomer ranging between 93 and 14 $\%$ (see Methods section).

\subsection{Characterization of the lattice location}

The VUV results, yielding a lower limit of 1-7\% of implanted ${}^{229\mathrm{m}}$Th nuclei decaying by photon emission, imply that at least several percent of the thorium atoms are incorporated with a local atomic configuration that favours the suppression of conversion-electron decay \cite{dessovic_229_2014}. To this end, emission channeling measurements using the radioactive decay radiation (see Methods section) are performed to characterize the thorium incorporation in the CaF$_2$ lattice. Since the lattice position of implanted thorium ions in CaF$_2$ cannot be determined using the ${}^{229}$Th isotope because of its long half-life of \qty{7920}{y}, $^{231}$Th (T$_{1/2}$= \qty{25.2}{h}) is used instead as a proxy. For that purpose, a VUV-grade CaF$_2$ crystal with a thickness of \qty{0.7}{\milli\meter} is implanted with a beam of mass $A=231$ at room temperature and with a fluence of \qty{1.0e12}{\centi\meter^{-2}}. Emission channeling patterns of $^{231}$Th are measured along four major crystal axes (Fig. \ref{fig2}) after a waiting period of \qty{4}{\hour} between the end of the implantation and the start of the measurement, to ensure that spectral contributions from the $\beta^-$decay of the shorter-lived parent isotopes ($^{231}$Fr, $^{231}$Ra and $^{231}$Ac) are negligible. 

If only one occupied site is considered, the best fit is obtained for thorium in a substitutional calcium site (Fig. \ref{fig2}). An occupation fraction of \qty{77\pm 4}{\%} is obtained from the fit. 
If additional high-symmetry sites for $^{231}$Th are considered, the quality of the fit does not improve, indicating that eventual fractions in such sites are below the detection limit (typically of the order of \qty{5}{\%}). The calcium substitutional fraction deduced from the fit needs to be corrected for secondary electrons from the sample and internal setup parts \cite{DAVIDBOSNE2020102}, leading to a final substitutional fraction between \qty{77}{\%} and \qty{100}{\%}. 

The root mean square (rms) displacement of the $^{231}$Th atoms with respect to the calcium site is also obtained from the fit, yielding \qty{0.20\pm 0.03}{\angstrom}. The fact that the rms displacement is significantly larger than the expected thermal vibration amplitude of a thorium atom in CaF$_2$ (\qty{0.08}{\angstrom}, calculated on the basis of the mass-defect model and the Debye temperature of CaF$_2$) indicates that additional defects, such as vacancies and self-interstitials, are present in the neighborhood of the thorium atoms. Such defects are formed in the collision cascades of the implanted ions \cite{pereira2019characterisation}. The trapping of such defects in the neighborhood of the thorium atoms may be enhanced by charge compensation mechanisms, for example, the energetically-favoured trapping of two interstitial fluorine atoms \cite{dessovic_229_2014}. Such charge-compensated configurations are predicted to preserve the large bandgap and thus suppress the conversion-electron decay of the ${}^{229\mathrm{m}}$Th isomer \cite{dessovic_229_2014}. 

\subsection{Discussion and conclusion}

The obtained excitation energy of the isomer is consistent with results from previous studies (see Figure \ref{fig:energy}). Our value is in agreement within the one $ \sigma$ confidence interval from conversion electron spectroscopy reported in \cite{seiferle_energy_2019} and two $\sigma$ confidence interval of the recent microcalorimetric measurement \cite{sikorsky_measurement_2020}, and decreases the uncertainty of the isomer's excitation energy by more than a factor of seven.

The measured half-life value of ${}^{229\mathrm{m}}$Th embedded in a MgF$_{2}$ crystal is lower compared to the calculated boundaries of the ${}^{229\mathrm{m}}$Th half-life \cite{kikunaga_half-life_2009,tkalya_radiative_2015,ruchowska_nuclear_2006,minkov_reduced_2017}, but should be considered as a lower limit because of the potential presence of non-radiative decay channels. Moreover, it should be corrected for the refractive index dependence of the crystal environment \cite{Tkalya_2000_refractiveindex, tkalya_radiative_2015}. Applying a $n^{3}$ dependence using the half-life value determined for $^{229\mathrm{m}}$Th embedded in a MgF$_{2}$ crystal and assuming a  refractive index of $n=1.488$ at \qty{148.7}{\nano\meter} a half-life of \qty{2.21(34)e3}{\second} can be inferred. Given the shallow implantation depth of the present samples, the validity of the ${n}^3$ dependence needs further investigation obtaining more precise half-life values and performing implantation under different conditions and in different crystals.

In conclusion, an alternative approach to populate and study the ${}^{229}$Th isomer via the beta decay of ${}^{229}$Ac incorporated in large-bandgap MgF$_{2}$ and CaF$_{2}$ crystals via implantation, allowed for the first time to observe the radiative decay of ${}^{229\mathrm{m}}$Th. A more precise energy value of the isomer was determined and the half-life of ${}^{229\mathrm{m}}$Th embedded in MgF$_{2}$ was reported. The observed incorporation of thorium in predominantly calcium substitutional sites, favouring the suppression of conversion-electron decay, is consistent with the VUV spectroscopy results that yield a conservative lower limit of 1-7 $\%$ for the fraction of ${}^{229\mathrm{m}}$Th that decays by photon emission. The observation of the radiative decay in a solid-state environment marks an important step towards the realization of a nuclear clock. Future studies combining emission channeling and VUV-spectroscopy, varying implantation and annealing parameters, and exploring different large-bandgap host crystals, will enable further optimization of this new approach. In particular, further development of the VUV-spectroscopy instrumentation will allow to decrease the uncertainty of the energy value by a factor of up to four and to study the isomer's half-life in different crystals, under different implantation conditions, providing valuable input in the search for laser excitation of the atomic nucleus. 

\bibliography{article}

\begin{figure}[hbt!]
	\centering
	\includegraphics[width=0.45\textwidth]{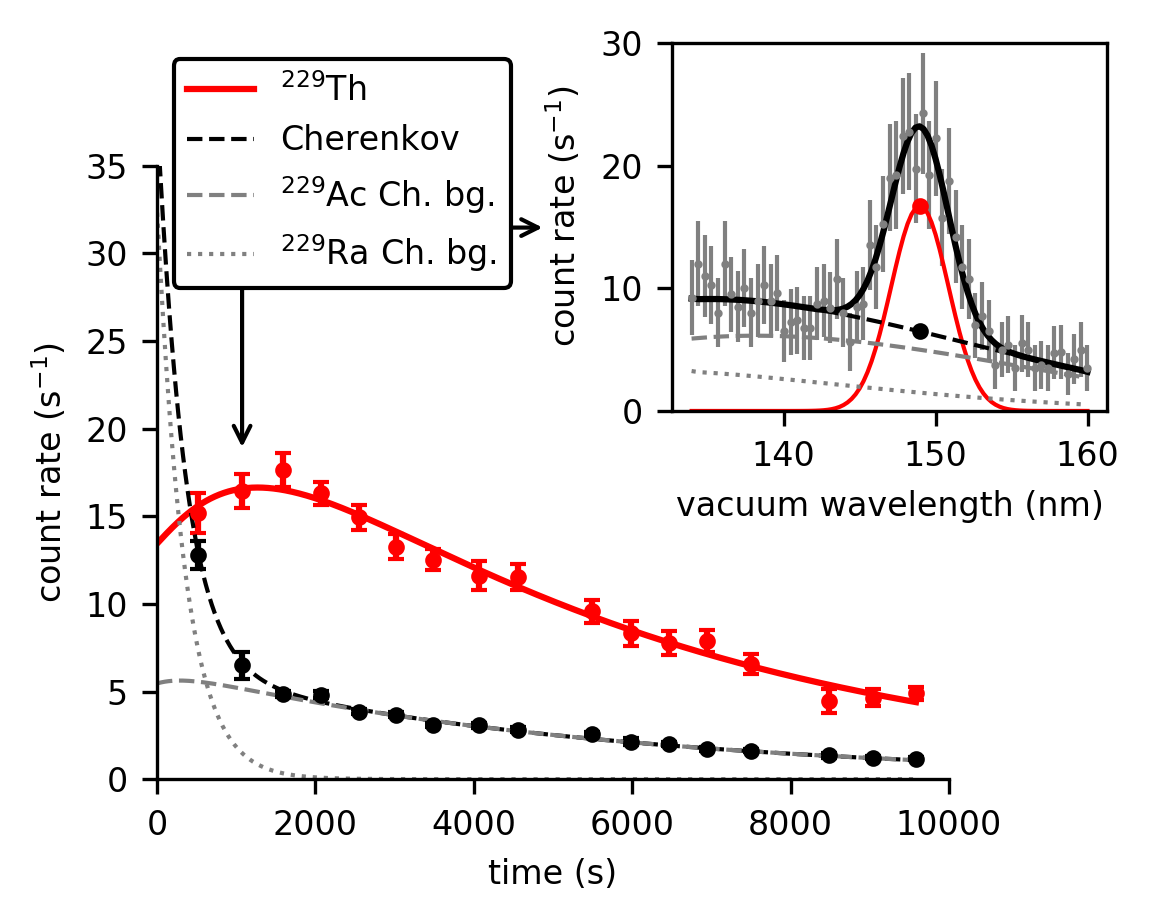}
	\caption{The fitted peak (red data points) and background (black data points) amplitude count rates of the VUV spectra as a function of time after implantation in the MgF${}_2$ - \qty{5}{\milli\meter} crystal are shown. Colored lines are the fit result of the data with a differential-equation model of the implantation and decay of the radioactive isotopes from which the half-life of $^{229\mathrm{m}}$Th is extracted.\\
	The inset shows the fit (solid black line) of the VUV wavelength spectrum measured around \qty{1073}{\second} after implantation to deduce the peak (red circle) and background (black circle) amplitudes. The data were collected using a \qty{2}{\milli\meter} spectrometer entrance slit width and \qty{5}{\second} measurement time per grating setting. The grey dotted and dashed lines represent the contribution of the Cherenkov radiation induced by the beta decay of ${}^{229}$Ra and ${}^{229}$Ac, respectively.  } 
	\label{fig:lifetime}
\end{figure}

\begin{figure}[hbt!]
	\centering
    \includegraphics[width=0.45\textwidth]{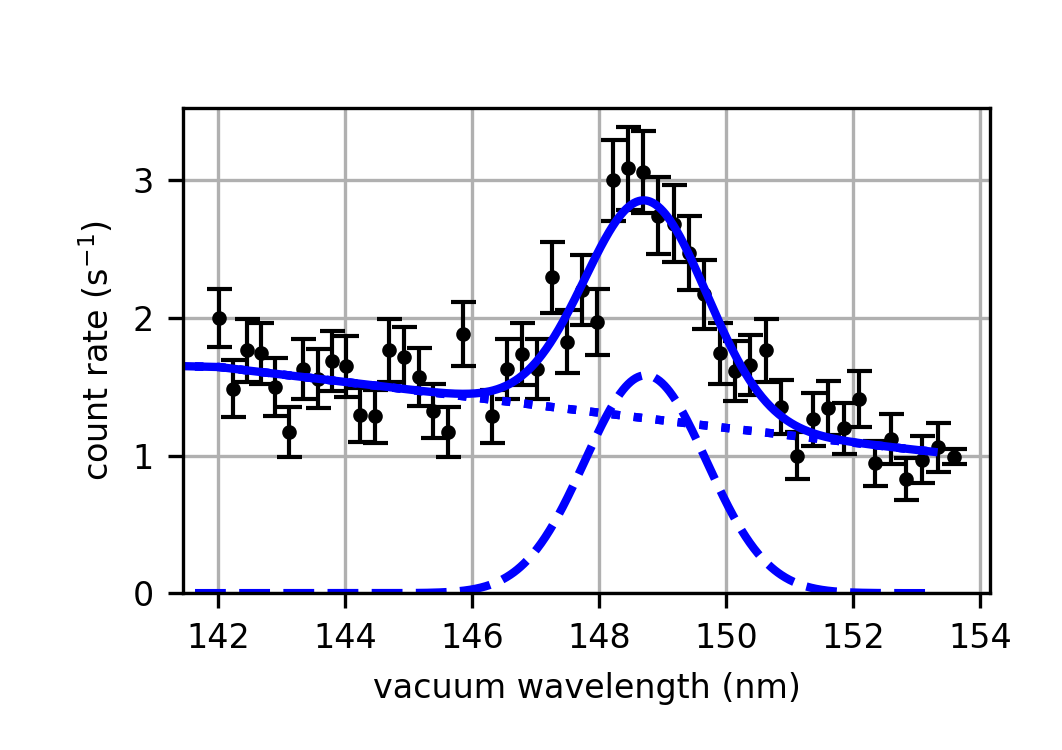}
	\caption{VUV spectrum taken with the CaF$_{2}$ - \qty{5}{mm} crystal and recorded with an entrance slit width of \qty{0.5}{\milli\meter}, along with a fit (solid line) of the data. Details on the fitting procedure are given in the Methods section. The dashed line shows the peak signal whereas the dotted line includes the constant dark count rate (0.3 s${}^{-1}$) and the Cherenkov background contributions. A resolution of \qty{2.25(21)}{\nano\meter} FWHM was obtained. The resulting wavelength value corresponds to the 11$^\mathrm{th}$ data value in Fig. \ref{fig:energy}. } 
\label{fig:narrowspectrum}
\end{figure}

\begin{figure}[hbt!]
	\centering
	\includegraphics[width=0.45\textwidth]{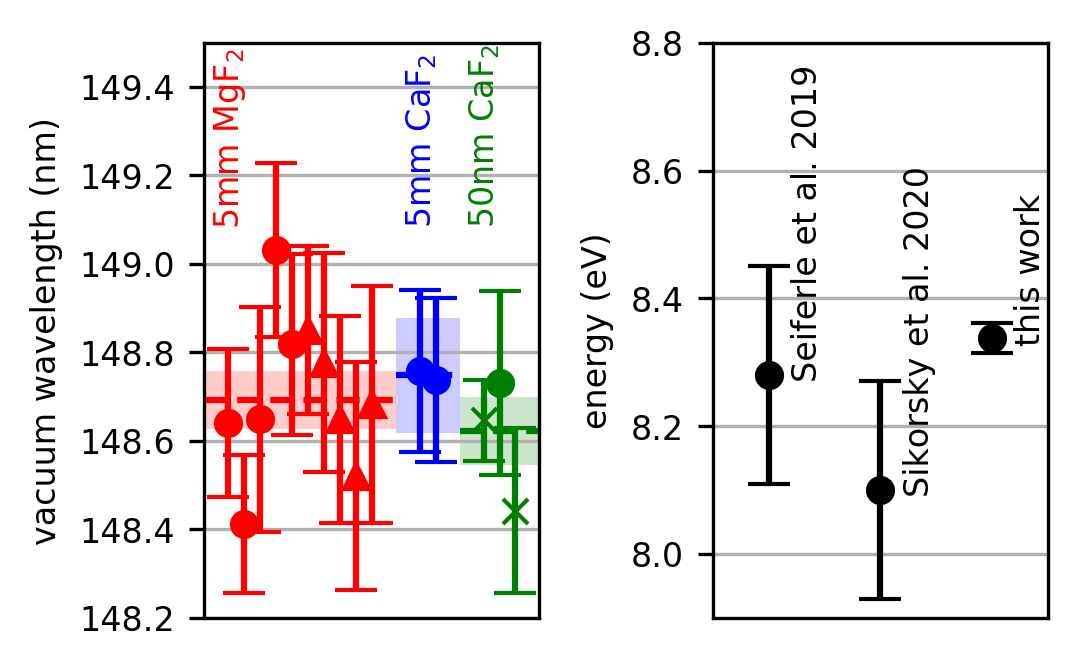}
	
	\caption{The left panel shows results from this work for single wavelength scans with small slit sizes (crosses: \qty{1}{\milli\meter}, circles: \qty{0.5}{\milli\meter}, triangles: \qty{0.25}{\milli\meter}) in three different crystals studied. Uncertainties are statistical and the weighted average and its uncertainty per crystal is represented by the dashed line and shaded area. The right panel compares the energy value deduced from the scans with slits \qty{\leq 0.5}{\milli\meter} in all three crystals, including the statistical and systematic uncertainties, to results of recent studies \cite{seiferle_energy_2019,sikorsky_measurement_2020}.}
	\label{fig:energy}
\end{figure}

\begin{figure}[h]
	\centering
	\includegraphics[width=0.45\textwidth]{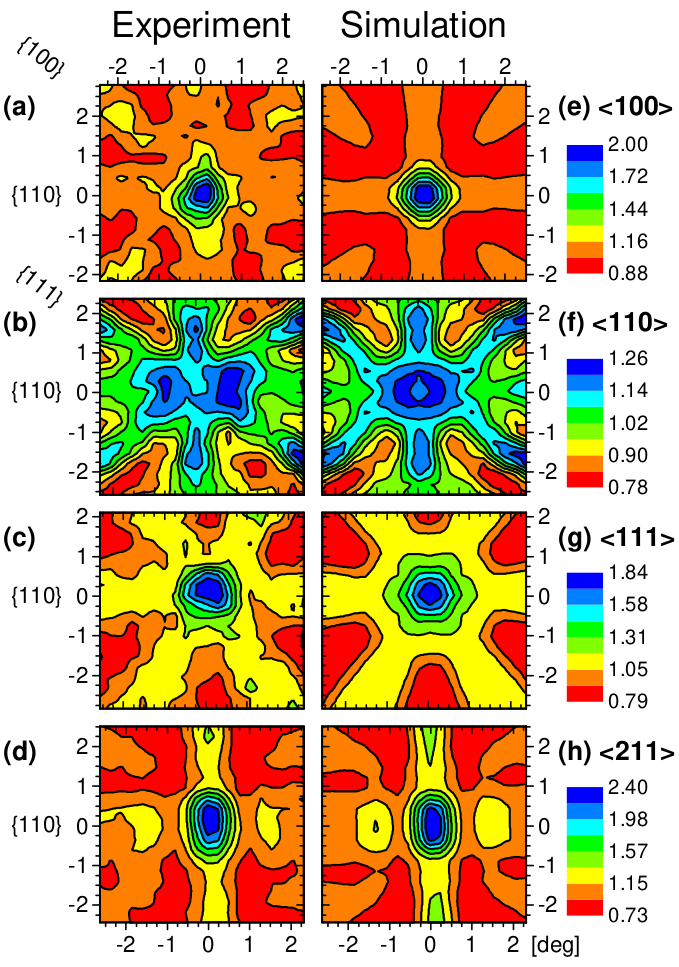}
	\caption{(a-d) Normalized experimental electron emission channeling patterns from $^{231}$Th in CaF$_2$ following room temperature implantation, in the vicinity of four major crystallographic axes: $\langle 100\rangle$, $\langle 110\rangle$, $\langle 111\rangle$ and $\langle 211\rangle$. (e-h) Best single-site fit with simulated patterns corresponding to calcium substitutional sites.}
	\label{fig2}
\end{figure}

\FloatBarrier

\section*{Methods}
\textbf{Radioactive ion beam production and implantation.}
Radioisotopes are produced by a 2\si{\micro\ampere} proton beam at 1.4~GeV impinging on a uranium carbide target. High temperatures (about \qty{2000}{\degreeCelsius}) lead to diffusion of the nuclear-reaction products towards the material boundary and subsequent effusion to the hot cavity ion source region, where surface ionization is used to create singly charged ions. The beam is accelerated to \qty{30}{\kilo\electronvolt} and mass separated  to obtain constituents of a specific isobar $A$ \cite{borge_isolde_2017}. The radioactive ion beam is implanted into one of the five crystals  mounted on the sample holder wheel. After implantation the wheel is turned by \qty{180}{\degree} in order to transfer the sample from its implantation position to the spectroscopy position facing the entrance slit of the vacuum-ultraviolet (VUV) spectrometer at a distance of about \qty{3}{\milli\meter}. 
From the gamma-ray spectra recorded during implantation, the implanted beam composition is deduced. Extended Data Table 1 summarizes the characteristics of the  $A=229$ and $A=230$ isobaric chain. During data collection with the \qty{50}{\nano\meter} thick CaF$_{2}$ crystal the production rate is a factor of 3 lower compared to the ones reported in the table. From $\beta$-decay studies, the total branching ratio feeding the isomer directly and indirectly in the $\beta$-decay of ${}^{229}$Ac has been determined between \qty{14} and \qty{93}{\%} \cite{ruchowska_nuclear_2006, verlinde_alternative_2019}.

\textbf{Large-bandgap crystals.}
Data on the large-bandgap crystals used in this study are summarized in the Extended Data Table \ref{tab:crystals}. The \qty{5}{\milli\meter} MgF$_{2}$ and \qty{5}{\milli\meter}, \qty{0.7}{\milli\meter} and \qty{0.5}{\milli\meter} CaF$_{2}$ are commercially available VUV grade crystals. The production procedure of the CaF$_{2}$ - 50 nm crystal is described below. From all crystals VUV photons from $^{229\mathrm{m}}$Th are observed. 

The CaF$_2$ thin film is grown using a Riber 49 Molecular Beam Epitaxy (MBE) reactor on a Si(111) substrate with a thickness of \qty{0.75}{\milli\meter}. The CaF$_2$ molecular beam is obtained by electron beam evaporation of a CaF$_2$ target. The flux is calibrated by using a quartz crystal microbalance, and the stability of the CaF$_2$ beam is ensured by a real time mass spectrometer partial pressure signal with feedback loop. A calibrated optical pyrometer is used to monitor the wafer surface temperature during the growth. Prior to loading into the MBE reactor, a 2 \% NH$_4$F wet clean of the Si(111) substrates is performed for 60 seconds to remove part of the native oxide and organic contamination. After introduction into the growth chamber the deoxidation step is carefully followed by reflection high-energy electron diffraction (RHEED) and the substrate is heated up to 750 $^{\circ}$C leading to the appearance of a strong $7\times 7$ surface reconstruction for the Si(111) substrate. In the present study, a CaF$_2$ thin film with a thickness of 50 nm is epitaxially grown at high temperature (740 $^{\circ}$C) with a growth rate of 3.5 nm/min. RHEED and atomic force microscopy (AFM) characterization of the CaF${}_2$ thin film is shown in Extended Data Figure \ref{figSICaF2}. The RHEED pattern along the [11-2] direction exhibits intense streaks which confirms the epitaxial property of the CaF$_2$ layer. The surface morphology of the CaF$_2$ layer is atomically smooth (root mean square roughness $< 4$ {\AA}) with the presence of steps and terraces, originating from the step-flow growth mode induced by the high growth temperature. 

\textbf{Emission channeling.}
The emission channeling technique is based on the interaction between the electrons emitted upon decay of a radioactive isotope and the periodic Coulomb potential of the host crystal, inducing channeling and blocking effects along major crystallographic axes and planes. The resulting angle-dependent emission yield is measured using a position-sensitive silicon pad detector \cite{wahl_position-sensitive_2004, silva_versatile_2013}, producing two-dimensional electron emission patterns. Since the emission anisotropy is strongly dependent on the position within the crystal lattice from where the electrons are emitted, it allows to identify which lattice sites are occupied by the implanted radioactive ions and to quantify the respective occupation fractions. The electron emission patterns are measured along various major crystallographic axes, by rotating the sample holder with a 3-axis goniometer. At least two axes are required for unambiguous lattice site identification; typically four axes are measured. Identification and quantification of the occupied lattice sites are obtained by numerically fitting measured and simulated emission patterns. The simulated patterns are calculated for various high-symmetry lattice sites using the Manybeam code \cite{hofsass_emission_1991}. Each measured pattern is then fitted with a linear combination of simulated sites \cite{DAVIDBOSNE2020102}, from where the occupation fractions are obtained. 

\textbf{Vacuum-ultraviolet spectroscopy.}
A Czerny-Turner mount grating monochromator with a \qty{100}{\milli\meter}$\times$\qty{100}{\milli\meter} plane grating and a sinusoidal ruling of 4000 lines per mm combined with a F/1.2 \qty{45}{\degree} off-axis parabolic collimation mirror with a focal length of \qty{119}{\milli\meter} is used. The device has a magnification of $M=1.5$ and the exit slit close to the detector is set accordingly. 
A Hamamatsu R8487 solar-blind photomultiplier tube detector with a quantum detection efficiency of \qty{19}{\percent} at \qty{150}{\nano\meter} is operated in single photon counting mode to detect diffracted photons passing the exit slit. 
The wavelength is set by rotation of the grating with a linear stepper motor driving a crankshaft and a linear wavelength calibration relation between motor position and the wavelength holds. The device is scanned from lower to higher wavelengths and the counts observed on the detector are recorded with a CAEN N6730 data acquisition unit. The grating drive offers a reproducibility of the grating position corresponding to a wavelength uncertainty of smaller than \qty{0.18}{\nano\meter} in the first diffraction order when scanned over a large spectral range.

A plasma light source operated with a nitrogen-oxygen gas mixture is mounted behind the crystal in spectroscopy position, emitting a collimated beam of light into the spectrometer with a wavelength spectrum consisting of several non-resolved multiplets of atomic transitions. 
An oxygen multiplet of transitions at \qty{130.2168}{\nano\meter}, \qty{130.4858}{\nano\meter}, \qty{130.6029}{\nano\meter} together with a multiplet of nitrogen transitions at \qty{149.2625}{\nano\meter}, \qty{149.2820}{\nano\meter}, \qty{149.4675}{\nano\meter} and a nitrogen multiplet at \qty{174.2731}{\nano\meter}, \qty{174.5249}{\nano\meter} are used for calibration. 
To validate the calibration, the single \qty{146.96}{\nano\meter} line in the spectrum of a cold krypton-xenon plasma light source (Excitech GmbH E-Lux) is used. The wavelength linewidth of the spectrometer is composed of a minimal achievable resolution for a close-to-diffraction-limited entrance slit settings (\SI{10}{\micro\meter}), dominated by the optical aberrations of the system, and an approximately linear increase of the linewidth with slit size. Experimentally, a full-width half-maximum linewidth of $\approx$\qty{2.5}{\nano\meter} and \qty{5.5}{\nano\meter} are observed for a \qty{0.5}{\milli\meter} and \qty{3}{\milli\meter} entrance slit, respectively.

Next to the contribution from the reproducibility of the grating position mentioned above, two more contributions to the systematic uncertainty of the wavelength calibration have been identified. (i) The calibration light source is positioned between \qty{217}{\milli\meter} and  \qty{350}{\milli\meter} from the entrance slit. 
The influence of an off-axis mounting of the source with a deviation of up to \qty{1}{\degree} is calculated and results in a systematic shift of the wavelength calibration of $<$ \qty{0.15}{\nano\meter}. (ii) The implanted radioactive beam profile and therefore the distribution of the light emission in the crystal remains poorly known. Typical radioactive beam spot sizes are of the order of a few mm (FWHM) in diameter. The systematic effect of different light source distributions on the observed wavelengths for entrance slit settings of $\gg$\qty{10}{\micro\meter} has been estimated by simulating the influence of the different light source distributions on the sample crystal for a number of entrance slit settings. Using slit settings $\leq$\qty{0.5}{\milli\meter}, the shift of the observed wavelength is less than \qty{0.33}{\nano\meter}.

The total systematic uncertainty for slit settings $\leq$\qty{0.5}{\milli\meter}, including the systematic effect of the wavelength calibration light source, the implantation beam profile and the grating drive reproducibility amounts in a conservative estimate to \qty{<0.41}{\nano\meter}.

The observed wavelength spectra around 148.7 nm consist of three components: (1) The detector dark count rate is constant over time and \qty{<1}{\per\second}, (2) the peak of the radiative decay of ${}^{229\mathrm{m}}$Th and (3) the continuous Cherenkov background induced by relativistic electrons from the radioactive decay of ${}^{229}$Ra and ${}^{229}$Ac. 
A typical spectrum is shown in Fig.~\ref{fig:narrowspectrum}. 
The rates from the Cherenkov background need to be scaled with the instantaneous activities of the corresponding fraction and have to be convoluted with the time behaviour of the decay chain elements.
The shape of the Cherenkov emission spectrum for the $\beta$-decay chains \mbox{${}^{230}$Ra $\rightarrow$ ${}^{230}$Ac $\rightarrow$ ${}^{230}$Th (I)} and \mbox{${}^{229}$Ac $\rightarrow$ ${}^{229}$Th (II)} is calculated using the Frank-Tamm-formula \cite{stellmer_radioluminescence_2015}. 
The obtained background spectrum from the $A=230$ decay chain (I) around \qty{149}{\nano\meter}
is due to the Cherenkov spectrum altered by the wavelength-dependent efficiency of the instrument (the wavelength bandpass) and by the time-dependence of the instantaneous $\beta$ activity at $A=230$. The Cherenkov spectrum altered by the wavelength bandpass can be described by a third order polynomial. The shape of the ${}^{229}$Ac background spectrum (II) is obtained by scaling the polynomial with the ratio of the calculated pure Cherenkov emission spectra and correcting with the instantaneous activity at $A=229$.

The $A=229$ spectra are fitted as a linear combination of the ${}^{229}$Ac Cherenkov background (altered by the wavelength bandpass) scaled with the instantaneous actinium activity, a Gaussian peak scaled with the instantaneous thorium isomer activity and a constant dark-count background. 
Figure \ref{fig:narrowspectrum} shows the background and the isomeric radiative decay peak  fitted to an observed spectrum. Narrow and broad linewidth scans used for the wavelength measurements are recorded sufficiently long after the end of implantation, such that the ${}^{229}$Ra Cherenkov background component vanishes, the ${}^{229\mathrm{m}}$Th activity is saturated and its time-dependence follows the simple exponential time-behaviour of the ${}^{229}$Ac activity.

\textbf{Extraction of the half-life.}
The spectrometer is continuously scanned from \SI{133}{\nano\meter} to \SI{160}{\nano\meter} with a \SI{2}{\milli\meter}-wide entrance slit setting and spectra are recorded. The amplitude of the observed photon peak and the contribution strengths of the radium and actinium Cherenkov emission are extracted from a fit (see Vacuum ultraviolet spectroscopy and c.f. inset of Figure \ref{fig:lifetime} for a sample fit).

The time dependence of the isomeric decay activity is fitted using a system of Bateman equations of the $A=229$ decay chain. The relative strength of the implantation beam constituents ${}^{229}$Fr, ${}^{229}$Ra and ${}^{229}$Ac and their half-lives are fixed to the results from gamma spectroscopy (see Extended Table 1) and the literature values, respectively. Next to the half-life of the isomer that is included as a free fitting parameter, a scaling factor is used to adjust the calculated isomer activity to describe the data. The scaling factor results from a product of the fraction of embedded ${}^{229\mathrm{m}}$Th nuclei that decays via radiative decay, the total efficiency of the VUV set-up and the total feeding probability of the isomer in the beta decay of ${}^{229}$Ac.   

Based on the simulated relative Cherenkov photon emission strength in the decay of ${}^{229}$Ra and ${}^{229}$Ac, the relative total Cherenkov photon emission rate is obtained from the activities of both decay chain elements after the end of implantation. This rate is scaled with a coefficient, that takes into account the efficiency of the VUV spectrometer for Cherenkov radiation. Figure \ref{fig:lifetime} shows good agreement between the observed background temporal behavior and the expected Cherenkov photon emission contributions.

\textbf{Ratio of radiative decaying versus embedded $^{229\mathrm{m}}$Th isotopes.} 
To estimate the ratio between the number of $^{229\mathrm{m}}$Th decaying via radiative decay versus the number of embedded $^{229\mathrm{m}}$Th after $\beta$-decay of $^{229}$Ac, several experimental parameters need to be considered of which for a number of them only estimates are available. A conservative lower limit is obtained using data from a \qty{7080}{s} long implantation in the CaF$_{2}$ - \qty{5}{\milli\meter} resulting in a VUV countrate of \qty{12}{\per\second} and \qty{1.2e6}{\becquerel} $^{229}$Ac source strength at \qty{5400}{\second} after the end of the implantation, combined with the following criteria. 

For a crystal implantation surface located at a distance of \qty{3}{\milli\meter} from the slit with a \qty{3}{\milli\meter} opening and a point-like light source at its center, the solid angle for photon collection is \qty{1.3}{\percent}. 
The first-order diffraction efficiency at \qty{150}{\nano\meter} of the grating used in this work is specified by the supplier to be \qty{40}{\percent} and the quantum detection efficiency of the detector \qty{19}{\percent}. 
A total detection efficiency of isotropically emitted photons from a point source at the center of the crystal implantation surface is $\approx$ \qty{0.1}{\percent} at 149 nm. Typically the radioactive ion beams from ISOLDE have beam profiles with an approximate full-width half-maximum of the intensity distribution of a few \qty{}{\milli\meter}, however, these conditions change during radioactive beam tuning and the optimization procedure applied in between the different measuring campaigns. Therefore, the overlap of the implanted beam profile with the region on the crystal implantation surface that is imaged through the \qty{3}{\milli\meter} $\times$ \qty{10}{\milli\meter} rectangular entrance slit on the spectrometer, is conservatively estimated to be $\leq$ \qty{100}{\percent}. The large-bandgap material leads to absorption of photons in the bulk and crossing the surface. Because of the small implantation depth, the VUV photons pass  a limited amount of bulk material and the surface facing the entrance slit. The transmission of the large-bandgap crystals are tested and transmission of VUV photons through the CaF$_{2}$ crystals vary between \qty{10} and \qty{80}{\percent} at \qty{149}{nm}. For the current estimate $<$\qty{100}{\percent} is assumed.

This leads to a lower limit of the ratio between the number of $^{229\mathrm{m}}$Th decaying via radiative decay versus the number of embedded $^{229\mathrm{m}}$Th after $\beta$-decay of $^{229}$Ac of $\geq$\qty{1}{\percent} or $\geq$\qty{7}{\percent} using the limits of the ${}^{229}$Ac $\beta$ branching ratio to ${}^{229\mathrm{m}}$Th values of \qty{93}{\percent} and \qty{14}{\percent}, respectively \cite{verlinde_alternative_2019}. Other implantation cycles with smaller slit settings or other crystals reveal consistent results.

\section{Acknowledgements}
We would like to thank the ISOLDE collaboration and technical group at CERN for their extensive support and assistance. This work has received funding from Research Foundation Flanders (FWO, Belgium), from GOA/2015/010 (BOF KU Leuven) and from  FWO and F.R.S.-FNRS under the Excellence of Science (EOS) programme (nr. 40007501), the Portuguese Foundation for Science and Technology (FCT, project CERN/FIS-TEC/0003/2019), the European Union's Horizon 2020 research and innovation programme under the ENSAR2 grant agreement no. 654002, under the Marie Skłodowska-Curie grant agreement no. 101026762 and the European Research Council (ERC) under the Thorium Nuclear Clock agreement no. 856415 and under the LRC agreement no. 819957.

\renewcommand{\figurename}{Extended Data Figure}
\setcounter{figure}{0} 
\renewcommand{\tablename}{Extended Data Table}
\setcounter{table}{0} 

\FloatBarrier
\newpage

\begin{table*}[h]
\begin{center}
\captionsetup{justification=centering}
\begin{minipage}{0.75\textwidth}
\renewcommand\footnoterule{}
\begin{tabular}{@{}lllll@{}}
\toprule
Isotope 
& Half-life & $\gamma$ energy (intensity${}^a$) & Q$_{\beta}$ value & Production rate\\
& & keV & keV & pps${}^b$\\
\midrule
$^{229}$Fr & 50.2(4) s & 336 (21.5 \%)
& 3126(21) & \num{1.6e5} \\
 & & 1256 (3.7 \%)
&  & \\
$^{229}$Ra & 4.0(2) min & 
& 1850(18) & \num{1.8e6}${}^c$ \\
$^{229}$Ac & 62.7(5) min & 569 (2.79 \%)
& 1111(12) &  $\leq$\num{e5}${}^d$\\
$^{229}$Th & 7932(28) y & 
& &  \\
& & &  & 
\\
$^{230}$Fr & 19.1(5) s & 706 (1.86 \%)
& 4983(32) &  \num{2.5e4} \\
 &  & 709 (16.8 \%)
&  & \\
$^{230}$Ra & 93(2) min & 253 (0.85 \%) & 678(19) &  \num{3.0e6} \\
&  & 285 (1.6 \%) 
&  & \\
$^{230}$Ac & 122(3) s & 1227 (1.0 \%)
& 2974(16) & $\leq$\num{e5} \\
 &  & 1244 (3.6 \%)
&  & \\
$^{230}$Th & 75400(300) y & 
& &  \\
$^{231}$Fr & 17.6(6) s & 
& 3864(14) &   \\
$^{231}$Ra & 104(1) s &  & 2454(17) &   \\
$^{231}$Ac & 7.5(1) min & 
&  1947(13)&  \\
$^{231}$Th & 25.52(1) h & 
& 391.5(15)&  \\
\botrule
\end{tabular}
\raggedleft
\footnotetext[1]{Absolute $\gamma$-ray intensity per $\beta$-decay}
\footnotetext[2]{Uncertainty on the production rate is estimated at 20 \%, implantation to implantation variation of the beam intensity rises up to a factor of 3}
\footnotetext[3]{$\gamma$ rays from $^{229}$Ra are too weak to observe, the production rate is determined using $^{229}$Ac $\gamma$ rays observed after the decay of $^{229}$Ra}
\footnotetext[4]{Lower limit deduced from the time evolution of the \qty{569}{\kilo\electronvolt} $\gamma$ ray in the decay of $^{229}$Ac}
\end{minipage}
\newline
\label{beams}
\caption{Characteristics of the isobaric  $A=229$~(\mbox{$^{229}$Fr $\rightarrow$ $^{229}$Ra $\rightarrow$ $^{229}$Ac $\rightarrow$ $^{229}$Th}), $A=230$~(\mbox{$^{230}$Fr $\rightarrow$ $^{230}$Ra $\rightarrow$ $^{230}$Ac $\rightarrow$ $^{230}$Th}) and \\ $A=231$~(\mbox{$^{231}$Fr $\rightarrow$ $^{231}$Ra $\rightarrow$ $^{231}$Ac $\rightarrow$ $^{231}$Th}) \\ $\beta$-decay chain. Values taken from Evaluated Nuclear Data Sheets \cite{browne_nuclear_2008}.}
\end{center}
\end{table*}

\begin{table*}[h]
\begin{center}
\begin{tabular}{@{}llll@{}}
\toprule
Material & Manufacturer  & Thickness \\
\midrule
MgF${}_2$ & Thorlabs Inc.& 5~mm  \\
CaF${}_2$ & Thorlabs Inc.& 5~mm  \\
CaF${}_2$ & MaTeck GmbH & 0.7~mm  \\
CaF${}_2$ & CRYSTAL GmbH & 0.5~mm  \\
CaF${}_2$ & (grown by the authors) & 50~nm \\
\botrule
\end{tabular}

\label{implantationcrystals}
\caption{Large-bandgap crystals used for VUV spectroscopy.}\label{tab:crystals}
\end{center}

\end{table*}

\begin{figure}[h]
	\centering
	\includegraphics[width=0.5\textwidth]{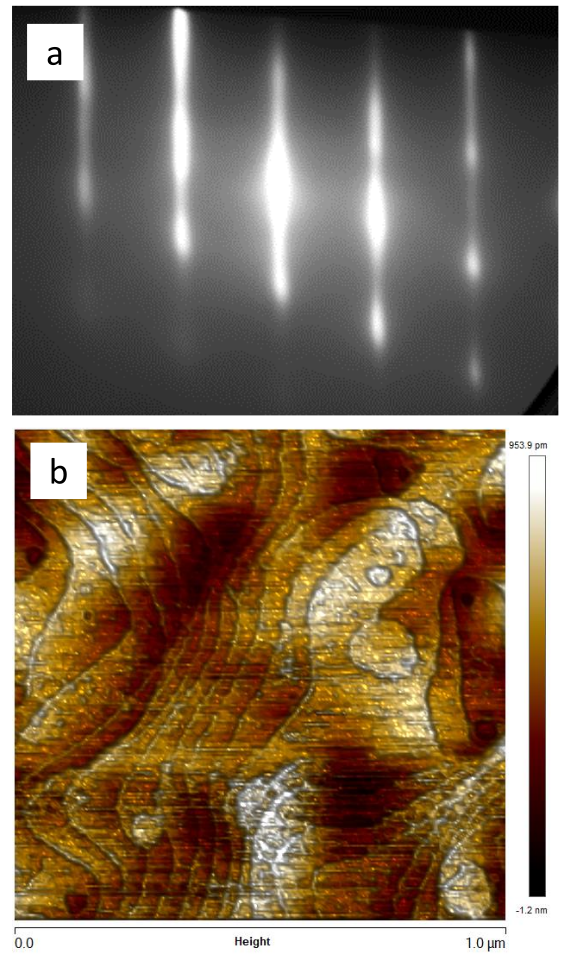}
	\caption{(a) RHEED pattern of the CaF$_2$ thin film along [11-2] azimuthal direction. (b) AFM image of the surface of the CaF$_2$ thin film.}
	\label{figSICaF2}
\end{figure}

\end{document}